\documentclass[11pt,a4paper,english,nofootinbib,,superscriptaddress]{revtex4}
\usepackage{lmodern}

\usepackage[T1]{fontenc}
\usepackage[latin9]{inputenc}
\usepackage{babel}
\usepackage{xcolor}
\usepackage{amsmath}
\usepackage{graphicx}
\usepackage{amssymb}
\usepackage{esint}
\usepackage[unicode=true, pdfusetitle,
 bookmarks=true,bookmarksnumbered=false,bookmarksopen=false,
 breaklinks=false,pdfborder={0 0 1},backref=false,colorlinks=false]
 {hyperref}
\def\b{\begin{equation}}
\def\e{\end{equation}}

\makeatletter
\@ifundefined{textcolor}{}
{%
 \definecolor{BLACK}{gray}{0}
 \definecolor{WHITE}{gray}{1}
 \definecolor{RED}{rgb}{1,0,0}
 \definecolor{GREEN}{rgb}{0,1,0}
 \definecolor{BLUE}{rgb}{0,0,1}
 \definecolor{CYAN}{cmyk}{1,0,0,0}
 \definecolor{MAGENTA}{cmyk}{0,1,0,0}
 \definecolor{YELLOW}{cmyk}{0,0,1,0}
 }

\usepackage{latexsym}\usepackage{bm}

\makeatother

\makeatother

\begin{document} 
 
\title{Lee-Wald Charge and Asymptotic Behaviors of the Weyl-invariant Topologically Massive Gravity}

\author{Suat Dengiz}

\email{sdengiz@thk.edu.tr}

\affiliation{Department of Mechanical Engineering, University of Turkish Aeronautical Association, 06790 Ankara, Turkey}

\author{Ercan Kilicarslan}

\email{ercan.kilicarslan@usak.edu.tr}

\affiliation{Department of Physics, Usak University, 64200, Usak, Turkey.}

\author{M. Reza Setare}

\email{rezakord@ipm.ir}

\affiliation{Department of Science, Campus of Bijar, University of Kurdistan, Bijar, Iran.}

\date{\today}
\begin{abstract}
  
We apply the Lee-Wald covariant phase space method to the Weyl-invariant Topologically Massive Gravity and compute the corresponding on-shell conserved charges. By using appropriate decay conditions for the existing propagating modes in the near-horizon of a stationary black hole, we obtain the charges generating the asymptotic symmetries. We show that the charges are integrable and the (modified) algebras among the asymptotic generators are \emph{closed} for the certain choice of central extensions. 

\end{abstract}
\maketitle

 \section{Introduction}
 As a Lagrangian field theory, general relativity describes the metric field as the dynamical quantity, and the associated Lagrangian is invariant under the local symmetry group of the diffeomorphisms belonging to the spacetime. Although the general relativity in general admits identical properties with the Standard model gauge theory in this perspective, they turn out to be strictly drifted apart in the Hamiltonian context. That is, unlike the ordinary Yang-Mills case, the Lie algebra of the Dirac constraints upon the phase space in general relativity during the canonical quantization becomes point-dependent yielding \emph{distinct} constraint functions in phase space. Hence, it fails to establish a viable group inducing the local diffeomorphisms of the Lagrangian \cite{Lee, Isham}. Lee and Wald have remarkably linked the configuration and phase spaces coherently. With this, they covariantly resolved the flaw between the local symmetries and constraints in 1990 by constructing the phase space symplectic form from the presymplectic one upon the configuration space generated via the integration of the symplectic current density over an initial Cauchy hypersurface \cite{Lee}. Using these fundamentals, then, they have achieved to construct the covariant Noether charge associated to any given local symmetry. Their method particularly turns out to be very successful in supplying the fundamentals of \emph{dynamical} black holes. To be more precise, for example, it is achieved to construct the \emph{local} conserved quantities at null infinity in the framework of Hamiltonian as the symplectic current radiates through the null infinity \cite{WaldZoupa} albeit the works on the BMS (Bondi-Metzner-Sachs) symmetry at that time \cite{BMS1, BMS2, BMS3, BMS4, BMS5, BMS6, BMS7, BMS8}. Herewith, the approach successfully reproduces the fundamental laws of the thermodynamics such as the Bekenstein-Hawking entropy for the dynamical black holes \cite{WaldBHenstropy, IyerWald}. It is also in a great consistency with the recent soft hairs (particles) model by Hawking, Perry and Strominger \cite{Strominger1, Strominger2, Strominger3, Strominger4} wherein an extended BMS symmetry disclosing the existence of infinite family of diffeomorphisms (namely, supertranslations) \cite{BarnichTroessaert1, BarnichTroessaert2, BarnichTroessaert3} is mainly addressed\footnote{As an example, see \cite{Prabhu} for the first law of black hole thermodynamics associated to the fields augmenting extra internal gauge degree of freedoms.}. Furthermore, the Lee-Wald (LW) covariant phase space method has been successfully applied to many diffeomorphism-invariant gravity theories possessing extra local symmetries such as $U(1)$ gauge symmetry. For example, see \cite{SetareAdami1, SetareAdami2} for the conserved charges and the entropy formulas associated to symmetry generators preserving near horizon conditions in the Einstein-Maxwell theory and Kerr-Newman (A)dS Black Holes, respectively. The details of how the (modified) LW covariant approach takes place will be presented in the bulk of the paper.

The failure of achieving unitarity and renormalizabilty simultaneously is the main problem in the construction of a complete $4$-dimensional quantum gravity. To get some insights into a possible well-behaved quantum gravity, countless number of works have been devoted to study the lower dimensional models. Particularly, studying cosmological Einstein's theory in $3$-dimensions has attracted huge attentions. Although the $3D$ cosmological Einstein gravity does not possess any local physical degree of freedom (dof), it has appealing global properties. For example, it possesses the Banados-Teitelboim-Zanelli (BTZ) black hole solutions \cite{BTZ} or $2D$ boundary CFT with the Brown-Henneaux central charge ($c=3l/2G$) dual to the bulk $AdS_3$ spacetime \cite{BrownHenneaux}. Here, the vital point is to construct a $3D$ gravity theory that also possess a local propagating dof as it has a viable boundary CFT. In this perspective, the Topologically Massive Gravity (TMG) is particularly interesting because it is a unitary and renormalizable theory and also describes a \emph{local dynamical} dof \cite{DJT}\footnote{Here the $AdS_3$ solution is dual to a $2D$ CFT admitting two copies of Virasoro algebra with the central charges $c_{L, R}=3l/2G  (\sigma \mp 1/k l)$, where $\Lambda= -1/l^2$; $L (R)$ is left(right) central charge and $k$ is the topological mass parameter.}. The cosmological TMG exhibits both gravitons and BTZ black hole solutions. We should mention that in this model with the usual sign for the gravitational constant, the massive excitations of cosmological TMG carry negative energy. In the absence of a cosmological constant, one can change the sign of the gravitational constant. But, if $\Lambda <0$, this will give a negative energy to the BTZ black hole, so the existence of a stable ground state is in doubt in this model. Moreover, TMG  has a bulk-boundary unitarity conflict. That is, either the bulk or the boundary theory is non-unitary, so there is a clash between the positivity of the two Brown-Henneaux boundary charges and the bulk energies. In the chiral gravity \cite{Chiral} which is a parity-odd theory of massive gravity in $2+1$ dimensions, the mass parameter takes a value  around AdS space such that either the left or right-moving central charge of the dual CFT vanishes. The mentioned problems of (cosmological) TMG yet remain  in such critical  points in the parameter space of models. The vanishing of one of the central charges is usually associated to the emergence in the bulk of a massless graviton mode, with a logarithmic fall-off behavior near the boundary. This log-mode has negative energy in the bulk, and it makes the dual CFT to be non-unitary.  Naively	one expects AdS to be the vacuum of the theory, but as it was shown \cite{AltasTekin} AdS may not be the true vacuum: in fact, it seems that one cannot do generic perturbations about it as the linearized metric is not determined by the linearized field equations alone. The true vacuum of the theory is currently unknown and certainly deserves a much more detailed study. This novel phenomenon requires to go to the second order in perturbation theory. An important issue about all 3D gravity theories is the unitarity of all perturbative and non perturbative excitations: as was shown in \cite{MaloneyStrominger, AltasTekin} in the case of chiral gravity, even if a perturbative (graviton) mode is found to be non-unitary, the background solution may turn out to be linearization unstable in the sense that the perturbative technique fails around the chosen background. Despite this ambiguous point of TMG, due to its unique properties towards a complete $2+1$-dimensional quantum gravity, it is an exceptional and foremost $2+1$ gravity theory.

 In this study, we further explore the \emph{particular extension} of TMG (namely, Weyl-gauged TMG) \cite{DengizErcanTekin} for the following reasons: with the help of an abelian gauge and a real scalar fields, the Weyl's gauging method replaces the rigid scale-invariance that necessitates the conformal flatness due to the Lorentz-invariance with a local scale-invariance leading to Poincar\'{e}-invariant field theories in generic curved spacetimes \cite{Weyl1, Weyl2, Weyl3, DengizTekin, TanhayiDengizTekin, Dengiz}. (Basics of the Weyl-invariance will be given in the next section). Using the fundamentals of the Weyl's approach, the Weyl-invariant extension of TMG is constructed in \cite{DengizErcanTekin}. Here it has been shown that the Weyl-gauged TMG unifies the ordinary TMG with Topologically Massive Electrodynamics and conformally coupled Proca mass term. The associated action does not possess any dimensional parameter: the local conformal symmetry is radiatively broken in the (A)dS spaces through the Coleman-Weinberg mechanism and spontaneously broken in the flat backgrounds \'{a} la Higgs mechanism. The breaking of Weyl's symmetry generates the masses of the excitations. Moreover, the model is shown to be unitary at the tree-level for certain regions of parameters. In this work, we specifically focus on the conserved charges and asymptotic symmetry group structures of the Weyl-gauged TMG. In this context, we apply the LW symplectic approach to the Weyl-gauged TMG \cite{DengizErcanTekin} and compute the associated quasi-local conserved charge. With the help of decay conditions for the propagating modes, we also study the near-horizon behaviors in details.

The lay-out of paper is as follows: In Sec.II, we briefly review the Topologically Massive Gravity and its Weyl-invariant extension. In Sec.III, we compute the field equations and the LW covariant charges of the Weyl-gauged TMG. In Sec.IV, we elaborately study the near-horizon behavior of the Weyl-gauged TMG via appropriate fall-off conditions for the existing dynamical fields. Sec.V is dedicated to our conclusions. 

\section{Topologically Massive Gravity and its Weyl-invariant extension}
\subsection{Cosmological Topologically Massive Gravity}
In this part, we shortly recapitulate the notable properties of cosmological topologically massive gravity (TMG) even if it is in general very well-known: as was mentioned above, the $2+1$-dimensional bare general relativity does not possess any physical degree of freedom (dof). This is still the case for the conformally-coupled scalar tensor theory\footnote{This can be easily seen during the transformation from Jordan to Einstein frames.}. In this respect, Deser, Jackiw and Templeton constructed an elegant $2+1$-dimensional unitary and renormalizable \emph{dynamical} gravity theory with the help of the gravitational Chern-Simons term in 1982 \cite{DJT}. The theory is called the (cosmological) TMG and described by the following action
 \begin{equation}
 S_{TMG}=\int d^3 x \sqrt{-g} \bigg [  m (\sigma R-2 \Lambda) +\frac{k}{2 } \epsilon^{\lambda \mu \nu} \Big ( \Gamma^\rho_{\lambda \sigma} \partial_\mu \Gamma^\sigma_{\nu \rho}+\frac{2}{3} \Gamma^\rho_{\lambda \sigma} \Gamma^\sigma_{\mu \tau}\Gamma^\tau_{\nu \rho}  \Big ) \bigg ].
 \label{tmg}
 \end{equation}
Here, $ m$ corresponds to the (dimensionful) Newton's constant and $ \epsilon^{\lambda \mu \nu} $ is a rank-3 tensor, while $ \sigma,k $ are dimensionless parameters. To have a unitary model in flat background, $\sigma$ is required to be negative. The Eq.(\ref{tmg}) describes a topologically massive graviton\footnote{With masses, $ M_{graviton}=-\frac{\sigma m}{\lvert k \rvert} $ and  $M^2_{graviton}=\frac{\sigma^2 m^2}{k^2 }+\Lambda $ about flat and (A)dS backgrounds, respectively \cite{Carlip}.} possessing a single helicity mode and acquires asymptotically AdS black hole solutions dubbed as BTZ black holes \cite{BTZ}. Despite its unique properties, TMG in general fails to be a complete theory in the holography context. More precisely, it has been shown that the energies of boundary gravitons and BTZ black holes contradict with each others as for generic mass parameter $ k$, and thus one of them inevitably turns out to be negative. 

\subsection{Weyl's gauge symmetry and its integration to TMG}
Since we elaborately study the Lee-Wald symplectic charges and asymptotic structure of Weyl-gauged TMG in the following sections, it is also useful to briefly remember the fundamentals of the Weyl-invariance and its integration to TMG: instead of the rigid scale-invariance\footnote{That is, $x^\mu \rightarrow \lambda x^\mu$ and $\Phi \rightarrow \lambda^d \Phi $ where $d$ is the scaling dimension of the field.} which demands the conformal flatness due to the Lorentz-invariance, the Weyl's gauge method considers a local scale-invariance in order to have Poincar\'{e}-invariant field theories in arbitrary curved spacetimes. Due to having the local conformal invariance, the Weyl-invariant field theories do not permit the presence of any dimensionful parameter \cite{Weyl1, Weyl2, Weyl3, DengizTekin, TanhayiDengizTekin, Dengiz}\footnote{Having the dimensionful parameter (i.e., Newton's constant) is the main problem of the non-renormalizibility of general relativity at the loop level. Therefore, since it is broken by the presence of any dimensionful parameter, the local-scale invariance has a great potential to be one of the core symmetries and so provide further inside in the quantum gravity.}. To see how the Weyl's local symmetry is achieved, let us first note that the action\footnote{See \cite{Ciambelli} for a recent study of Weyl's symmetry in the flat-holography perspective and \cite{Scholz} for a comprehensive review of Weyl's gauging approach.}
\begin{equation}
S_{\Phi}=- \frac{1}{2}\int d^3 x \sqrt{-g} \, g^{\mu \nu} \partial_\mu \Phi \, \partial_\nu \Phi
\label{scalarkinetic}
\end{equation}
fails to be invariant under the ensuing Weyl's conformal transformations in $2+1$ dimensions 
\begin{equation}
g_{\mu\nu} \rightarrow g^{'}_{\mu\nu}=e^{2 \zeta(x)} g_{\mu\nu}, \hskip 1 cm \Phi \rightarrow \Phi^{'} =e^{-\frac{\zeta(x)}{2}} \Phi.
\label{gphitrns}
\end{equation}
Here,  $\zeta(x)$ is a function of $x$. By replacing the usual partial derivative with the gauge covariant derivative 
\begin{equation}
{\cal{D}}_\mu \Phi =\partial_\mu\Phi -\frac{1}{2} A_\mu \Phi, \hskip 1 cm {\cal{D}}_\mu g_{\alpha \beta}=\partial_\mu g_{\alpha\beta}+ 2 A_\mu g_{\alpha \beta},
\end{equation}
wherein $A_\mu$ is the so-called abelian Weyl's gauge field and satisfies $A_\mu \rightarrow A^{'}_\mu = A_\mu - \partial_\mu \zeta(x)$, one has 
\begin{equation}
 \tilde{D}_\mu g_{\alpha \beta} \rightarrow ( \tilde{D}_\mu g_{\alpha \beta})^{'}=e^{2 \sigma(x)} \tilde{D}_\mu g_{\alpha \beta} , \hskip 1 cm \tilde{D}_\mu \Phi \rightarrow  (\tilde{D}_\mu \Phi)^{'}=e^{-\frac{\sigma(x)}{2}} \tilde{D}_\mu \Phi ,
\end{equation}
that clearly leads to the Weyl-invariant extension of Eq.(\ref{scalarkinetic}) \cite{DengizTekin, TanhayiDengizTekin, Dengiz}. Similarly, the bare Maxwell action also fails to be invariant under the transformations in Eq.(\ref{gphitrns}) which necessitates an extra scalar field with a specific order to be Weyl-invariant as follows
\begin{equation}
S_{E_\mu} =  - \frac{1}{2} \int d^3 x \sqrt{-g}\,\, \Phi^{-2} F_{\mu \nu} F^{\mu \nu}.
\label{weylmaxwell} 
\end{equation}
To integrate the Weyl's gauge symmetry to gravity sector, one needs to define the following Weyl-invariant connection 
 \begin{equation}
 \widetilde{\Gamma}^\lambda_{\mu\nu}=\frac{1}{2}g^{\lambda\sigma} \Big ({\cal{D}}_\mu g_{\sigma\nu}+{\cal{D}}_\nu g_{\mu\sigma}-{\cal{D}}_\sigma g_{\mu\nu} \Big).
 \label{weylconnection}
 \end{equation}
Using Eq.(\ref{weylconnection}), one gets the Weyl-invariant entensions of the Riemann tensor as follows
\begin{equation}
\begin{aligned}
\tilde{R}^\mu{_{\nu\rho\sigma}}&=\partial_\rho \tilde{\Gamma}^\mu_{\nu\sigma}-\partial_\sigma \tilde{\Gamma}^\mu_{\nu\rho}
+ \tilde{\Gamma}^\mu_{\lambda\rho} \tilde{\Gamma}^\lambda_{\nu\sigma}-\tilde{\Gamma}^\mu_{\lambda\sigma} \tilde{\Gamma}^\lambda_{\nu\rho} \\
& =R^\mu{_{\nu\rho\sigma}}+\delta^\mu{_\nu}F_{\rho\sigma}+2 \delta^\mu{_[\sigma} \nabla_{\rho]} A_\nu 
+2 g_{\nu[\rho}\nabla_{\sigma]} A^\mu \\
& \hskip 3.2 cm+2 A_[\sigma   \delta_{\rho]}\,^\mu A_\nu + 2 g_{\nu[\sigma}  A_{\rho]} A^\mu  +2 g_{\nu[\rho} \delta_{\sigma]}\,^\mu  A^2 , 
\label{weylriem}
\end{aligned} 
\end{equation}
with $ 2 A_{[ \rho} A_{\sigma]} \equiv A_\rho A_\sigma -  A_\sigma A_\rho$ and $A^2= A_\mu A^\mu$ \cite{DengizTekin, TanhayiDengizTekin, Dengiz}. With Eq.(\ref{weylriem}), the Weyl-invariant extension of the Ricci tensor becomes
\begin{equation}
\tilde{R}_{\nu\sigma} = \tilde{R}^\mu{_{\nu\mu\sigma}}=R_{\nu\sigma}+F_{\nu\sigma}-\Big [\nabla_\sigma A_\nu - A_\nu A_\sigma +A^2  g_{\nu\sigma} \Big ]-  g_{\nu\sigma}\nabla \cdot A,
\end{equation}
where $\nabla \cdot A \equiv \nabla_\mu  A^\mu$. Finally, the Weyl-gauged Ricci scalar becomes
\begin{equation}
\tilde{R}=R-4\nabla \cdot A-2 A^2,
\label{weyltransricciscalar}
\end{equation}
 transforming as $\tilde{R} \rightarrow (\tilde{R})^{'} = e^{-2 \zeta (x) } \tilde{R}$ and thus is not Weyl-invariant. As in the Maxwell case, one gets the Weyl-invariant extension of $3$-dimensional Einstein's gravity by virtue of extra scalar field as follows  \cite{DengizTekin, Dengiz}
  \begin{equation}
S= \int d^3 x \sqrt{-g} \, \Phi^2 \, \tilde{R}= \int d^n x \sqrt{-g} \, \Phi^2 \, \Big (R-4\nabla \cdot A-2 A^2 \Big).
\label{weyleinstein}
\end{equation}

As for the core model studied in the current paper, by following the procedure shortly described above (and particularly using the Weyl-invariant connection in Eq.(\ref{weylconnection}), the Weyl-invariant enhancement of TMG has been constructed in \cite{DengizErcanTekin}: 
\begin{equation}
 \begin{aligned}
 S_{Weyl-TMG}=&\int d^3x  \sqrt{-g}\,\, \sigma \Phi^2 [R-4 \nabla . A-2 A^2] \\
 &+ \frac{k}{2}\int d^3 x \sqrt{-g} \,\, \epsilon^{\lambda \mu \nu} \Big ( \tilde{\Gamma}^\rho_{\lambda \sigma} \partial_\mu \tilde{\Gamma}^\sigma_{\nu \rho}+\frac{2}{3} \tilde{\Gamma}^\rho_{\lambda \sigma} \tilde{\Gamma}^\sigma_{\mu \tau} \tilde{\Gamma}^\tau_{\nu \rho}  \bigg ),
 \label{witmg} 
 \end{aligned}
 \end{equation}
whose explicit form is given in Eq.(\ref{witmg3}) to avoid writing it twice. Referring \cite{DengizErcanTekin} for the details, one should observe that the Weyl's conformal symmetry is much more universal than the ordinary conformal symmetry: as is mentioned ensuingly, even though the abelian Chern-Simons term cannot come about in the conformally-invariant form of TMG, it intriguingly emerges in the Weyl-gauged TMG. As is seen in the explicit form of the action in Eq.(\ref{witmg3}), the Weyl-TMG interestingly assembles the usual TMG, the Topologically Massive Electrodynamics and a conformally-coupled Proca mass term \cite{DengizErcanTekin}. Moreover, after a straightforward calculations, one can show that the relationship between the Lagrangian of Weyl-gauged and ordinary Chern-Simons terms becomes 
\begin{equation}
 \begin{aligned}
 {\cal L}_{CS(\widetilde{\Gamma})}={\cal L}_{CS(\Gamma)}+ \frac{k}{4}\epsilon^{\lambda \mu \nu} A_\lambda F_{\mu\nu}  -\partial_\mu \Big [ \frac{k}{2}\epsilon^{\lambda \mu \nu} g^{\alpha \sigma} (\partial_\lambda g_{\nu \sigma}) A_\alpha- \frac{k}{2} \epsilon^{\lambda \mu \nu} \Gamma^\rho_{\lambda \rho} A_\nu  \Big ].
 \end{aligned}
 \end{equation}
Also, together with the following Weyl-invariant scalar and Maxwell sectors 
 \begin{equation}
 S_{\Phi}=-\frac{\alpha}{2} \int d^3x \sqrt{-g}\,\,({\cal D}_\mu \Phi {\cal D}^\mu \Phi+\nu \Phi^6) \,, \hskip 0.5cm \hskip 1cm S_{A^\mu}=-\frac{\beta}{4}\int d^3x \sqrt{-g}\,\, \Phi^{-2} F_{\mu\nu}F^{\mu \nu},
 \end{equation}
 where $\alpha,\beta $ are dimensionless quantities, it has been shown that the Weyl-gauged TMG is unitary at the tree-level around the constant curvature backgrounds for certain values of the parameters. Here, as in the \cite{DengizTekin, TanhayiDengizTekin, Dengiz}\footnote{See \cite{JackiwPi} for a similar symmetry breaking mechanism by Jackiw and Pi.}, the Weyl's symmetry is spontaneously broken by the virtue of ordinary Higgs and Coleman-Weinberg mechanisms in (A)dS and flat vacua, respectively \cite{DengizErcanTekin}. Observe that only the Chern-Simons term without any physical dof persists to exist in the limit of symmetric vacuum (i.e., $ \langle \Phi \rangle=0 $). The gauge field must disappear because of the Maxwell term. Generally speaking, as one considers higher derivative terms in gauge or gravity sectors in the Weyl-gauging method, the symmetric and non-symmetric vacua turn out to be detached. That is inevitable in the existence of higher derivative terms. However, one can demand that it is essential for the symmetry to be broken as in the Standard Model Higgs mechanism, and thus the symmetric vacuum is not allowed, whereas the model reduces to the ordinary TMG, the Topologically Massive Electrodynamics and a conformally-coupled Proca mass term in its broken phase (i.e., $ \langle \Phi \rangle=\sqrt{m} $) \cite{DengizErcanTekin}. 

\section{Lee-Wald Charges of Weyl-gauged TMG}
Now that we have gone over the essential preliminaries, let us now compute the quasi-local charges for the Weyl-gauged TMG via the LW covariant phase space approach. To do so, let us first notice that the explicit form of the Lagrangian Eq.(\ref{witmg}) becomes
 \begin{equation}
\begin{aligned}
L [\psi]&= \sqrt{-\mbox{g}} \bigg \{ \sigma \Phi^2 [R-4 \nabla \cdot A-2 A^2]+ \frac{k}{2} \epsilon^{\lambda \mu \nu} \bigg ( \Gamma^\rho_{\lambda \sigma} \partial_\mu \Gamma^\sigma_{\nu \rho}+\frac{2}{3} \Gamma^\rho_{\lambda \sigma} \Gamma^\sigma_{\mu \tau} \Gamma^\tau_{\nu \rho}  \bigg )\\
&\hskip 1.2 cm + \frac{k}{4} (\epsilon^{\lambda \mu \nu} A_\lambda F_{\mu\nu}-\frac{\beta}{k}\Phi^{-2} F_{\mu\nu}F^{\mu \nu}) -\frac{\alpha}{2} ({\cal D}_\mu \Phi {\cal D}^\mu \Phi+\nu \Phi^6)\\
&\hskip 1.2 cm-\partial_\mu \Big [ \frac{k}{2}\epsilon^{\lambda \mu \nu} g^{\alpha \sigma} (\partial_\lambda g_{\nu \sigma}) A_\alpha- \frac{k}{2} \epsilon^{\lambda \mu \nu} \Gamma^\rho_{\lambda \rho} A_\nu  \Big ] \bigg \},
\label{witmg3} 
\end{aligned}
\end{equation}
where we keep the dimensionless parameter for the sake of future discussions. At that step, one should notice that since the usual LW symplectic method is only valid for the covariant theories and the Chern-Simons term breaks the covariance of Lagrangian, a naive direct attempt of using LW approach to get the charge for Weyl-gauged TMG will not work. For this reason, we will follow the modified LW approach proposed by Tachikawa \cite{Tachikawa} to handle the Lagrangian involving Chern-Simons term. (See also \cite{Kim, SetareAdami3, SetareAdami4} for the related works)\footnote{In our calculation, since the Lagrangian of the Weyl-TMG contains extra dofs together with the graviton field as in \cite{SetareAdami5} where the LW charges in the Einstein-Maxwell-Dilation theory are obtained, we will follow it, too.}. After underlying this crucial point and denoting the dynamical fields as  $ \psi=\{g_{\mu\nu}, A_\mu, \Phi\} $ as a compact form, one gets the first order variation of the Eq.(\ref{witmg3}) as follows
\begin{equation}
\delta L[\psi]=E_{\psi}[\psi]\delta \psi+\partial_\mu \Theta^\mu[\psi, \delta \psi].
\label{firstordvart}
\end{equation}
As is well-known, setting $E_\psi$ to zero gives the on-shell field equations for the propagating fields. Namely, by doing so, one gets the metric field equation as 
\begin{equation}\label{2}
\begin{aligned}
E_{(g)}^{\mu \nu} = - \sqrt{-\mbox{g}} \biggl\{ & \sigma \biggl[ \Phi^{2} G^{\mu \nu} + g^{\mu \nu} \Box \Phi^{2} - \nabla^{\mu} \nabla^{\nu} \Phi^{2} - 2 \Phi^{2} A^{\mu} A^{\nu} \\
& +g^{\mu \nu} \Phi^{2} A^{2} + 4 A^{(\mu} \nabla^{\nu )} \Phi^{2} - 2 g^{\mu \nu} A \cdot \nabla \Phi^{2} \biggr] \\
& + k C^{\mu \nu} - \frac{\alpha}{2} \left[ {\cal D}^{\mu} \Phi {\cal D}^{\nu} \Phi -\frac{1}{2} g^{\mu \nu} \left( {\cal D}_{\alpha} \Phi {\cal D}^{\alpha} \Phi +\nu \Phi^{6} \right) \right] \\
& -\frac{\beta}{2} \Phi^{-2} \left[ F^{\mu}_{\hspace{1.5 mm} \lambda} F^{\nu \lambda} -\frac{1}{4} g^{\mu \nu} F_{\alpha \beta}F^{\alpha \beta} \right] \biggr\}=0.
\end{aligned}
\end{equation}
Here, $G^{\mu\nu}=R^{\mu\nu}-\frac{1}{2}g^{\mu\nu}R$ is the ordinary Einstein's tensor. Similarly, the Weyl's gauge field equation turns out to be 
\begin{equation}\label{3}
E_{(A)}^{\mu} = \sqrt{-\mbox{g}} \left[ \left( 4 \sigma +\frac{\alpha}{4}\right) {\cal D}^{\mu} \Phi^{2} + \frac{k}{2} \epsilon^{\lambda \mu \nu} F_{\nu \lambda} - \beta \nabla_{\nu} \left( \Phi^{-2} F^{\mu \nu}\right) \right]=0,
\end{equation}
where $D_{\mu} \Phi^{2}= \partial_{\mu} \Phi^{2} - A_{\mu} \Phi^{2}$ due to the fact that one has $\Phi^{2} \rightarrow (\Phi^{2})^{\prime} = e^{-\zeta} \Phi^{2}$ under the local Weyl's conformal transformation. Furthermore, the scalar field equation becomes
\begin{equation}\label{4}
\begin{aligned}
E_{\Phi}= \sqrt{-\mbox{g}} \biggl[ & 2 \sigma \Phi \left( R-4 \nabla \cdot A - 2 A^{2} \right) + \alpha \left( \Box \Phi -\frac{1}{2} \Phi \nabla \cdot A -\frac{1}{4}\Phi A^{2}-3 \nu \Phi^{5} \right) + \frac{\beta}{2} \Phi^{-3} F_{\alpha \beta}F^{\alpha \beta}  \biggr]=0.
\end{aligned}
\end{equation}
Finally, the boundary term which is called "symplectic potential" will read
\begin{equation}
\begin{aligned}
\Theta^{\mu}[\psi ; \delta \psi] = \sqrt{-\mbox{g}} \biggl\{ & \sigma \biggl[ 2 \Phi^{2} \nabla^{[\alpha} \left( g^{\mu ] \beta} \delta g_{\alpha \beta} \right) - 4 \Phi^{2} g^{\mu \nu} \delta A_{\nu} +4 \Phi^{2} g^{\mu \nu} \delta g_{\nu \lambda} A^{\lambda} \\
& -2 \Phi^{2} g^{\alpha \beta} \delta g_{\alpha \beta} A^{\mu} - 2 g^{\alpha [ \mu} \nabla^{\beta ]} \Phi^{2} \delta g_{\alpha \beta} \biggr] \\
& + \frac{k}{2} \epsilon^{\lambda \mu \nu} \left( \Gamma^{\alpha}_{\lambda \beta} \delta \Gamma^{\beta}_{\nu \alpha} - 2 R^{\alpha}_{\lambda} \delta g_{\alpha \nu} \right) +\frac{k}{2} \epsilon^{\lambda \mu \nu} A_{\lambda} \delta A_{\nu} \\
& -\alpha D^{\mu} \Phi \delta \Phi -\beta \Phi^{-2} F^{\mu \nu} \delta A_{\nu}\biggr\}.
\label{symppot}
\end{aligned}
\end{equation}

Denoting the generators of the diffeomorphism and $U(1)$ gauge symmetry respectively as $\xi^\mu(x)$ and $\lambda (x)$, one can define the incorporated symmetry generator as $\chi=(\xi, \lambda)$ under which one has the following infinitesimal transformations of the dynamical dofs \cite{SetareAdami1, SetareAdami2}
\begin{equation}
	\delta_\chi g_{\mu\nu}=\mathsterling_\xi g_{\mu\nu}, \qquad \delta_\chi A_\mu=\mathsterling_\xi A_\mu+\partial_\mu \lambda, \qquad \delta_\chi \Phi =\mathsterling_\xi \Phi, 
\end{equation}
where $\mathsterling_\xi$ is the Lie derivative with respect to the vector field $\xi$. Then, by following the modified LW symplectic approach in \cite{Tachikawa}, one can demonstrate that the variation of the Lagrangian in Eq.(\ref{witmg3}) generated by the enhanced generator $\chi$ can be recast as follows
\begin{equation}
\delta_\chi L[\psi]= \mathsterling_\xi L[\psi]+\partial_\mu  \Xi^\mu_\chi [\psi],
\label{liederlag}
\end{equation}
where 
\begin{equation}
\Xi^\mu_\chi [\psi]=\frac{k}{2}\sqrt{-\mbox{g}} \epsilon^{\lambda\mu\nu} \Big[\partial_\nu \Gamma^\beta_{\lambda\alpha} \partial_\beta \xi^\alpha+\Gamma^\alpha_{\lambda\alpha}\partial_\nu \lambda-\partial_\lambda g_{\alpha\nu}\partial^\alpha \lambda+A_\lambda \partial_\nu \lambda \Big].
\end{equation}
Observe that since the Lagrangian fails to be invariant under the enlarged symmetry transformations generated by $\chi=(\xi, \lambda)$, the symplectic potential (i.e., surface term) in Eq.(\ref{symppot}) will not turn out to be a covariant quantity in that case. Rather, as one varies the symplectic potential with respect to the combined transformation \'{a} la $\chi$, one finds that it actually deviates from its Lie derivative as follows \cite{Tachikawa, Kim, SetareAdami3, SetareAdami4}
\begin{equation}
\delta_\chi \Theta^{\mu}[\psi ; \delta \psi]= \mathsterling_\xi \Theta^{\mu}[\psi ; \delta \psi]+ \Pi^\mu_\chi[\psi ; \delta \psi],
\label{vartheta}
\end{equation}
where
\begin{equation}
\begin{aligned}
\Pi^\mu_\chi[\psi ; \delta \psi]&=\frac{k}{2}\sqrt{-\mbox{g}} \epsilon^{\lambda\mu\nu} [ \partial_\nu \delta \Gamma^\beta_{\lambda\alpha} \partial_\beta \xi^\alpha+\delta \Gamma^\alpha_{\lambda\alpha} \partial_\nu \lambda-\partial_\lambda \delta g_{\alpha\nu}\partial^\alpha \lambda-\partial_\lambda g_{\alpha \nu} \delta g^{\alpha\beta} \partial_\beta \lambda+\delta A_\lambda \partial_\nu \lambda]    \\
&+\partial_\nu \Big\{-\frac{k}{2}\sqrt{-\mbox{g}} \epsilon^{\lambda\mu\nu} [\delta\Gamma^\beta_{\lambda\alpha} \partial_\beta \xi^\alpha+2 \delta \Gamma^\alpha_{\lambda \alpha} \lambda+2 \delta g_{\alpha \lambda} \partial^\alpha \lambda] \Big\}.
\end{aligned}
\end{equation}
Moreover, taking the variation of Eq.(\ref{firstordvart}) as the one induced by the enlarged symmetry generator $\chi$ and accordingly addressing the Eq.(\ref{liederlag}), one will be able to define the on-shell Noether current as follows
\begin{equation}
J^\mu_\chi [\psi]=\Theta^{\mu}[\psi ; \delta_\chi \psi] -\xi^\mu L[\psi]-\Xi^\mu_\chi [\psi],
\label{onshellcurrent}
\end{equation}
in such a way that one gets  $\partial_\mu J^\mu_\chi [\psi] \simeq 0$ where the symbol $"\simeq"$ stands for the on-shell equality. Subsequently, with the help of Poincare lemma, one can show that the on-shell current in Eq.(\ref{onshellcurrent}) can be rewritten as 
\begin{equation}
J^\mu_\chi [\psi] \simeq \partial_\nu K^{\mu\nu}_\chi [\psi],
\end{equation}
where
\begin{equation}
\begin{aligned}
 K^{\mu\nu}_\chi [\psi] &=\sqrt{-\mbox{g}} \Big \{ -2 \sigma \Big (\Phi^2 \nabla^{[\mu}\xi^{\nu]}+2 \xi^{[\mu}\nabla^{\nu]} \Phi^2-4 \Phi^2 \xi^{[\mu} A^{\nu]}  \Big) \\
& \hskip 1.2 cm -2 k \epsilon^{\lambda\mu\nu} \Big(S_{\lambda\alpha} \xi^\alpha-\frac{1}{4} \Gamma^\alpha_{\lambda\beta} \nabla_\alpha \xi^\beta  \Big)+\frac{k}{2} \epsilon^{\lambda\mu\nu} A_\lambda (A_\alpha \xi^\alpha -2 \lambda)-\beta \Phi^{-2} F^{\mu\nu}(A_\alpha \xi^\alpha - \lambda)\Big\}.
\label{pncarelemma}
\end{aligned}
\end{equation}
Here, the Schouten tensor is $S_{\mu\nu}=R_{\mu\nu}-\frac{1}{4} g_{\mu\nu}R$. Besides, by varying the Eq.(\ref{onshellcurrent}) and then making use of Eq.(\ref{vartheta}), one finds that the LW symplectic potential turns out to be 
\begin{equation}
\begin{aligned}
w^\mu_{\mbox{LW}} [\psi; \delta\psi, \delta_\chi \psi]&=\frac{1}{16 \pi} \Big( \delta \Theta^{\mu}[\psi ; \delta_\chi \psi]-\delta_\chi \Theta^{\mu}[\psi ; \delta \psi]-\Theta^{\mu}[\psi ; \delta_{\delta \chi} \psi]\Big)\\
& \simeq \frac{1}{16 \pi} \partial_\nu \Big (\delta K^{\mu\nu}_\chi [\psi]-K^{\mu\nu}_{\delta \chi} [\psi]+2 \xi^{[\mu}  \Theta^{\nu]}[\psi ; \delta \psi] \Big)\\
& + \frac{1}{16 \pi} \Big (\delta  \Xi^\mu_\chi [\psi]- \Xi^\mu_{\delta \chi} [\psi]- \Pi^\mu_\chi[\psi ; \delta \psi] \Big).
\label{symplecticform}
\end{aligned}
\end{equation}
Notice that after a straightforward calculation, one can show that the terms in the second parenthesis in Eq.(\ref{symplecticform}) can be recast as a total derivative as follows
\begin{equation}
\delta  \Xi^\mu_\chi [\psi]- \Xi^\mu_{\delta \chi} [\psi]- \Pi^\mu_\chi[\psi ; \delta \psi]  = \partial_\nu \Sigma^{\mu\nu}_\chi  [\psi ; \delta \psi],
\label{etxtraterm}
\end{equation}
where 
\begin{equation}
\Sigma^{\mu\nu}_\chi  [\psi ; \delta \psi]=\frac{k}{2}\sqrt{-\mbox{g}} \epsilon^{\lambda\mu\nu} \Big (\delta \Gamma^\beta_{\lambda\alpha}\partial_\beta \xi^\alpha+2 \delta \Gamma^\alpha_{\lambda\alpha} \lambda+2 \delta g_{\alpha\lambda} \partial^\alpha \lambda \Big).
\label{extrabts}
\end{equation}
Followingly, by substituting Eq.(\ref{etxtraterm}) into the expression for the LW symplectic current in Eq.(\ref{symplecticform}), one will arrive at 
\begin{equation}
w^\mu_{\mbox{LW}} [\psi; \delta\psi, \delta_\chi \psi] = \partial_\nu {\cal Q}^{\mu\nu}_\chi [\psi; \delta \psi],
\end{equation}
where the formula for the LW charge (symplectic 2-form) turns out to be
\begin{equation}
{\cal Q}^{\mu\nu}_\chi [\psi; \delta \psi]=\frac{1}{16 \pi} \Big\{\delta K^{\mu\nu}_\chi [\psi]-K^{\mu\nu}_{\delta \chi} [\psi]+2 \xi^{[\mu}  \Theta^{\nu]}[\psi ; \delta \psi] + \Sigma^{\mu\nu}_\chi  [\psi ; \delta \psi] \Big\}.
\label{lwcharge}
\end{equation}
Ultimately, by inserting Eq.(\ref{symppot}), Eq.(\ref{pncarelemma}) and Eq.(\ref{extrabts}) into Eq.(\ref{lwcharge}), one can show that the on-shell LW charge for the Weyl-invariant TMG becomes\footnote{Observe that for the appropriate choices of the dynamical fields, Eq.(\ref{chargeddss}) recovers the symplectic charge for the TMG obtained in \cite{Nazaroglu}.}
\begin{equation}
\begin{aligned}
16 \pi {\cal Q}^{\mu\nu}_\chi [\psi; \delta \psi]&=-2 \sigma \sqrt{-\mbox{g}} \bigg \{\Phi^2 \Big(\frac{1}{2} h \nabla^{[\mu}\xi^{\nu]}- h^{\alpha[\mu} \nabla_\alpha \xi^{\nu]}+\xi^\alpha \nabla^{[\mu} h^{\nu]}_\alpha-\xi^{[\mu} \nabla_\alpha h^{\nu]\alpha}+ \xi^{[\mu}\nabla^{\nu]}h \Big)\\
& \hskip 2 cm +\delta \Phi^2 \Big ( \nabla^{[\mu}\xi^{\nu]}- 2 \xi^{[\mu} A^{\nu]} \Big)+2 \xi^{[\mu} D^{\nu]} \delta \Phi^2-\xi^{[\mu} h^{\nu]\alpha} D_\alpha \Phi^2-\Phi^2 \xi^{[\mu} h^{\nu]\alpha} A_\alpha \bigg \}\\
&-2 k \sqrt{-\mbox{g}} \epsilon^{\mu\nu\rho} \bigg \{\delta S_{\rho \alpha} \xi^\alpha-\frac{1}{2} \delta \Gamma^\alpha_{\rho\beta}\nabla_\alpha \xi^\beta+\xi^\beta R^\alpha_{[\beta}h_{\rho]\alpha}-\frac{1}{4} \lambda \partial_\rho h-\frac{1}{2} h_{\alpha\rho}\partial^\alpha \lambda \bigg\}\\
& +k  \sqrt{-\mbox{g}} \epsilon^{\mu\nu\rho} \delta A_\rho \Big (A_\alpha \xi^\alpha -\lambda \Big)-2 \alpha \sqrt{-\mbox{g}}\, \xi^{[\mu} D^{\nu]} \Phi \delta \Phi \\
&-\beta \sqrt{-\mbox{g}} \Phi^{-2} \bigg \{ \Big[ \Big( \delta F^{\mu\nu}+\frac{1}{2}h F^{\mu\nu} \Big)-2 \Phi^{-1} \delta \Phi F^{\mu\nu} \Big](A_\alpha \xi^\alpha-\lambda) +F^{\mu\nu}\delta A_\alpha \xi^\alpha+ 2 \xi^{[\mu}F^{\nu]\alpha} \delta A_\alpha \bigg\},
\label{chargeddss}
\end{aligned}
\end{equation}
where we have made use of the following definitions
\begin{equation}
h_{\mu\nu} =\delta g_{\mu\nu}; \hskip 1 cm \delta g^{\mu\nu}=-h^{\mu\nu}; \hskip 1 cm h=h^\alpha_\alpha=g^{\alpha\beta} h_{\alpha\beta}.
\end{equation}

\section{Fall-off of Fields and Asymptotic Charges}
Now we wish to study the near horizon behavior of a stationary black hole in the Weyl-invariant TMG in order to particularly reveal the behavior of the Weyl's gauge symmetry and check if there comes any residual symmetry in this region of the spacetime. For this purpose, let us assume that the geometry of the near horizon associated with a stationary black hole is described by the following generic metric in the Gaussian null coordinates as in \cite{Tamburino, Giribet1, Giribet2, SetareAdami1, SetareAdami2, SetareAdami5}
\begin{equation}
g=-2 \kappa (x) \rho dv^2 +2 dv d\rho+2 \rho \theta (x)dv dx+\Big(\Omega^2(x)+\rho \hat{\lambda} (x)\Big) dx^2.
\label{nearhrznmtrc}
\end{equation}
Here, $\kappa$ is the surface gravity and $\nu$ is the so-called advanced coordinate which provides to describe any null surface via $g^{\mu\nu} \partial_\mu \nu \partial_\nu \nu =0$. Moreover, the generator of the surface is $k^\mu=g^{\mu\nu}\partial_\nu \nu$ and affinely parameterized by $\rho$. As is apparent in the Eq.(\ref{nearhrznmtrc}), we start with the general situation where all functions $\kappa, \theta, \Omega, \hat{\lambda}$ are assumed to be $x$-dependent \cite{Tamburino, Giribet1, Giribet2, SetareAdami1, SetareAdami2, SetareAdami5}. By picking the gauge-fixing condition to be $A_\rho=0$, one is allowed to set up the ensuing decays of the Weyl's real scalar and gauge fields up to ${\cal O}(\rho^2)$ in the vicinity of horizon, respectively
\begin{equation}
\Phi= \Phi^{(0)}(x)+\rho \Phi^{(1)}(x)+{\cal O}(\rho^2) ,
\label{gaugefldper} 
\end{equation}
\begin{equation}
A_\nu=A^{(0)}_\nu (x) + \rho A^{(1)}_\nu (x)+{\cal O}(\rho^2), \hskip 0.5 cm A_x = A^{(0)}_x (x) + \rho A^{(1)}_x (x)+{\cal O}(\rho^2),
\label{gaugefieldper}
\end{equation}
where we initially suppose that all the components depend on $x$ again. Note that with the Eq.(\ref{gaugefieldper}), one gets the non-vanishing components of field-strength tensor as
\begin{equation}
F_{\nu\rho}=-A^{(1)}_\nu-2 \rho A^{(2)}_\nu+{\cal O}(\rho^2), \hskip 0.5 cm F_{\nu x}=-\partial_x A^{(0)}_\nu-\rho \partial_x A^{(1)}_\nu+{\cal O}(\rho^2), \hskip 0.5 cm F_{\rho x}=A^{(1)}_x+2 \rho A^{(2)}_x+{\cal O}(\rho^2).
\label{fldstrngthper}
\end{equation}
Then, by using the Eq.(\ref{nearhrznmtrc}), Eq.(\ref{gaugefldper}), Eq.(\ref{gaugefieldper}) and Eq.(\ref{fldstrngthper}) in the field equations obtained in the previous section, one will arrive at the constrains that the surface gravity $\kappa$ and $\Phi_0$ must be constant and $A^{(0)}_v=0$. 

To get the asymptotic generators for the enhanced symmetry, we first presume that the boundary conditions are \emph{state-independent} which is achieved by taking the leading terms to be independent of the propagating modes as in \cite{Giribet1, SetareAdami5}. It is actually straightforward to show that by solving
\begin{equation}
\mathsterling_\chi g_{\rho\rho}=0, \qquad \mathsterling_\chi g_{\nu\rho}=1, \qquad \mathsterling_\chi g_{\rho\rho}=0,
\end{equation}
one ends up with
\begin{equation}
\begin{aligned}
&\xi^\nu=T(x)+{\cal O}(\rho^3), \hskip 0.7 cm \xi^{\rho}=\frac{1}{2\Omega^2(x)} \theta(x) T^{'}(x) \rho^2 +{\cal O}(\rho^3), \\
&\xi^x=Y(x)-\frac{1}{\Omega^2(x)} T^{'}(x) \rho+\frac{\hat{\lambda}}{2\Omega^4(x)} T^{'} \rho^2+ {\cal O}(\rho^3), \\
& \lambda=\lambda^{(0)}+\frac{1}{\Omega^2(x)} A^{(0)}_x (x) T^{'}(x) \rho-\frac{1}{2\Omega^4(x)}  [\hat{\lambda} A^{(0)}_x  -\Omega^2(x) A^{(1)}_x ]T^{'}\rho^2+{\cal O}(\rho^3),
\label{asympttcklvctrs}
\end{aligned}
\end{equation}
where $(')$ denotes derivative with respect to $x$ and $T, Y, \lambda^{(0)}$ represent any functions depending on $x$. Later, with the help of Eq.(\ref{asympttcklvctrs}) and $\delta_\xi g_{\mu\nu}=2 \nabla _{(\mu}\xi_{\nu)}$, one eventually gets the  variation of the propagating fields with respect to the enlarged symmetry generator as follows \cite{Giribet1, SetareAdami5}
\begin{equation}
\begin{aligned}
\delta_\chi \theta &=\mathsterling_Y \theta - 2 \kappa \partial T, \hskip 0.5 cm \delta_\chi \Omega =\mathsterling_Y \Omega, \hskip 0.5 cm \delta_\chi \kappa =0, \hskip 0.5 cm  \delta_\chi \hat{\lambda} = \mathsterling_Y \lambda +2 \theta \partial_x T-2 \partial^2_x T\\
 \delta_\chi A^{(0)}_\nu &=0, \hskip 0.5 cm \delta_\chi A^{(1)}_\nu =\mathsterling_Y A^{(1)}_\nu, \hskip 0.5 cm \delta_\chi A^{(0)}_x =\mathsterling_Y A^{(0)}_x+A^{(0)}_\nu \partial_x T+\partial_x \lambda^{(0)}\\
 & \hskip 3 cm \delta_\chi A^{(1)}_x =\mathsterling_Y A^{(1)}_x+A^{(1)}_\nu \partial_x T.
\label{variatoffieldss}
\end{aligned}
\end{equation}
Observe that the gauge-parameters are field-dependent. Therefore, before computing the algebra among the asymptotic charges, let us first search for the algebra among the generators: as is well-known, the algebra among the generators in general will not be closed in the field-dependent case. To cure this obstacle, one needs to define an appropriate modified Lie derivative to check for the closeness of algebra via $[\delta_{\xi_1}, \delta_{\xi_2}]g_{\mu\nu}=\delta_{[\xi_1, \xi_2]_M}g_{\mu\nu}$ yielding \cite{Giribet1, Giribet2, BarnichTroessaert1, BarnichTroessaert2, BarnichTroessaert3, SetareAdami5, SheikhJabbariCompere}\footnote{See also \cite{LambertPHD} for a related comprehensive thesis.}
\begin{equation}
\left[\xi_1(T_1, Y_1), \xi_2(T_2, Y_2)\right]_{*}=\mathsterling_{\xi_1}  \xi_2-\delta^{(g)}_{\xi_1} \xi_2+\delta^{(g)}_{\xi_2} \xi_1.
\label{mdfbrack}
\end{equation}
Here, $\delta^{(g)}_{\xi_1} \xi_2 $ represents the correction coming to the $\xi_2$ because of the variation of the metric via $\xi_1$. At this moment, it is worth mentioning an important point about the uniqueness of the modification: observe that the modification we follow originates from the physical reason that the vectors are field-dependent. Applying the standard Lie brackets and considering the variation of the field-dependency make this modification unique in order to absorb this field-dependency \cite{Giribet1, Giribet2, BarnichTroessaert1, BarnichTroessaert2, BarnichTroessaert3, SheikhJabbariCompere}\footnote{This can also be seen as one applies the Leibniz rule for derivatives of the field-dependent vectors.}. But it does not mean that there is no other alternative modification which may close the algebra. Now, using the variations of fields in Eq.(\ref{variatoffieldss}) and also the short-hand notation $\delta^{(g)}_{\xi_1} \xi_2=\xi^\mu_2(\delta_{\xi_1}f_i)$ where $f_i$ stands for the fields in the metric (such as $\Omega^2$ etc.), one then has 
\begin{equation}
\begin{aligned}
&\xi^\nu_2(\delta_{\xi_1}f_i)=0, \qquad \xi^\rho_2(\delta_{\xi_1}f_i)=\frac{1}{2\Omega^4(x)}  \Big [ (\mathsterling_{Y_1} \Omega(x)) \theta(x) T^{'}_2+\Omega^2(x) \Big(\mathsterling_{Y_1} \theta(x)-2 \kappa T^{'}_1 \Big) T^{'}_2 \Big ] \rho^2, \\
& \xi^x_2(\delta_{\xi_1}f_i)=-\frac{1}{\Omega^4(x)}(\mathsterling_{Y_1} \Omega(x)) T^{'}_2 \rho+\frac{\hat{\lambda}}{\Omega^6(x)} (\mathsterling_{Y_1} \Omega(x)) T^{'}_2 \rho^2\\
& \hskip 5.6 cm  +\frac{1}{2\Omega^4(x)} \rho^2 \Big [\mathsterling_{Y_1} \lambda(x)+2\theta(x) T^{'}_1-2 \bar{\nabla}_C \bar{\nabla}_D T_1\Big] T^{'}_2. 
\end{aligned}
\end{equation}
All in all, by making use of these setups and after a straightforward calculations, one can demonstrate that the enhanced algebra associated to the diffeomorphism sector becomes closed as 
\begin{equation}
\left[\xi_1 (T_{1}, Y_{1}), \xi_2 (T_{2}, Y_{2})\right]_{*}=\xi_{12} (T_{12}, Y_{12}),
\end{equation}
where the functions in $\xi_{12}$ are
\begin{equation}
T_{12}=Y_1 \partial_x T_2-Y_2 \partial_x T_1, \quad Y_{12}=Y_1 \partial_x Y_2-Y_2 \partial_x Y_1.
\label{fncmdf}
\end{equation}
Moreover, note that the extended symmetry $\chi=\chi(T, Y, \lambda^{(0)})$ contains the extra mode $\lambda^{(0)}(x)$ associated to the $U(1)$ abelian gauge symmetry. As was shown in the \cite{SetareAdami5} which is our case, too, with the help of the definition of the extra commutations 
\begin{equation}
\begin{aligned}
&[\chi(0, 0, \lambda^{(0)}_1), \chi(0, 0, \lambda^{(0)}_2)]_{*} =0, \\
[\chi(0, 0, \lambda^{(0)}_1), \chi(0, Y_2, 0)]_{*}&=- [\chi(0, Y_2, 0), \chi(0, 0, \lambda^{(0)}_1)]_{*}=\chi(0, 0, - \mathsterling_{Y_2}\lambda^{(0)}_1),
\end{aligned}
\end{equation}
the algebra among the asymptotic generators turns out to be closed
\begin{equation}
[\chi_1, \chi_2]=\chi_{12},
\end{equation}
where $\lambda^{(0)}_{12}= \mathsterling_{Y_1}\lambda^{(0)}_2- \mathsterling_{Y_2}\lambda^{(0)}_1$. Moreover, as was demonstrated in \cite{SetareAdami5}, by setting the modes as
\begin{equation}
\begin{aligned}
T_{(m, n)}&=\chi(z^m \bar{z}^n, 0, 0), \qquad \quad Y_m=\chi(0, -z^{m+1}, 0) \\
\bar{Y}_m &=\chi(0, -\bar{z}^{m+1}, 0),\quad \lambda^{(0)}_{(m, n)}=\chi(0, 0, z^m \bar{z}^n) \quad ;\quad m, n \in \mathbb{Z},
\label{suprtrswitt}
\end{aligned}
\end{equation}
where $z$ and $\bar{z}$ are complex coordinates on the conformal two-sphere, one gets the following family of brackets associated to the modes
\begin{equation}
\begin{aligned}
\left[Y_m, Y_n\right]&=(m-n) Y_{m+n},\quad \left[\bar{Y}_m, \bar{Y}_n\right]=(m-n) \bar{Y}_{m+n}, \quad \left[Y_m, \bar{Y}_n\right]=0\\
\left[T_{(m, n)}, T_{(r, p)}\right]&=0, \quad \left[Y_r, T_{(m, n)}\right]=-m T_{(m+r, n)}, \quad \left[\bar{Y}_r, T_{(m, n)}\right]=-n T_{(m, n+r)}\\
\left[\lambda^{(0)}_{(m, n)}, \lambda^{(0)}_{(r, p)}\right]&=0, \quad \left[Y_r, \lambda^{(0)}_{(m, n)}\right]=-m \lambda^{(0)}_{(m+r, n)}, \quad \left[\bar{Y}_r, \lambda^{(0)}_{(m, n)}\right]=-n \lambda^{(0)}_{(m, n+r)},\quad \left[\lambda^{(0)}_{(m, n)}, T^{(0)}_{(r, p)}\right]=0,
\end{aligned}
\end{equation}
which describes a family of supertranslation $(T_{(m, n)})$, two family of Witt algebras $(Y_m\,\mbox{and}\,\bar{Y}_m)$ and a family of multiple charges $(\lambda^{(0)}_{(m, n)})$ \cite{SetareAdami5}.

Now that we have obtained the modified brackets among the enhanced symmetry generators associated to the diffeomorphism plus $U(1)$ gauge symmetry, we can now study the integrability and algebra among the asymptotic charges. For this purpose, let us first recall that the variation (fluctuation) of the charges can be defined by integrating the LW symplectic charge throughout the horizon as\footnote{Since the existing symmetry structure is same as the one in \cite{SetareAdami5}, we closely follow this paper in our analysis.}
\begin{equation}
\delta Q_\chi= \oint_{Horizon} dS^{Horizon}_{\mu\nu} {\cal Q}^{\mu\nu}_{LW} [\psi, \delta \psi; \chi].
\end{equation}
Furthermore, as in \cite{SetareAdami5}, one can define the asymptotic charge associated to the enhanced symmetry generator $\chi=\chi(\xi, \lambda)$ as a one-parameter phase space integration as follows
\begin{equation}
Q_\chi= \int_{0}^{1}ds \oint_{Horizon} dS^{Horizon}_{\mu\nu} {\cal Q}^{\mu\nu}_{LW} [\psi; \chi\, \lvert\, s].
\end{equation}
Here, $s$ is the parameter describing the curve in the phase space. Noting that the background contribution is generally described by $s=0$ whose subtraction will yield a finite charge. After taking this point into account, let us also notice that the LW algebra among charges is given in terms of the Dirac brackets as \cite{SetareAdami5}
\begin{equation}
\{Q_{\chi_1}, Q_{\chi_2}\}_{LW}= Q_{[\chi_1, \chi_2]}+\hat{{\cal C}}(\chi_1, \chi_2),
\label{LWalgebra}
\end{equation} 
where $\hat{{\cal C}}(\chi_1, \chi_2)$ stands for the central extension. Moreover, the algebra in the Eq.(\ref{LWalgebra}) can be recast as follows
\begin{equation}
\{Q_{\chi_1}, Q_{\chi_2}\}_{LW}=\delta_{\chi_2} Q_{\chi_1}.
\label{LWalgebra2}
\end{equation}
By making use of all these settings in Eq.(27), after a long and somewhat cumbersome calculation, one will eventually end up with that the asymptotic charge \emph{in general} fails to be integrable. However, this result interestingly turns into the desired form for specific choices of gauge components. That is, the associated integral can be written as a \emph{total derivative} only for two distinct choices, namely $A^{(0)}_x=0$ or $A^{(1)}_x=0$. For the first choice, one gets $\nu\rho$-component of the asymptotic charge as
\begin{equation}
\begin{aligned}
\hat{Q} (\chi)&=\frac{1}{16\pi} \int_{\mbox{H}} dS\, \sqrt{\Omega} \bigg \{\sigma \Phi^2_0 \Big[2\kappa T-Y \theta \Big]+k \bigg [Y\frac{(2\kappa\hat{\lambda}+\theta^2-2 \theta^{'} )}{2\sqrt{\Omega}}+Y \frac{\theta \Omega^{'}}{2 \Omega^{3/2}}+Y^{'} \frac{\Omega^{'}}{2 \Omega^{3/2}}\\
&\hskip 3 cm +Y \frac{\Omega^{'2}}{8 \Omega^{5/2}}+\lambda^{(0)}\frac{\Omega^{'}}{2 \Omega^{3/2}}-T \frac{\theta}{\sqrt{\Omega}} \bigg]+\beta \Phi^{-2}_0 \lambda^{(0)} A^{(1)}_\nu  \bigg\} .
\end{aligned}
\end{equation}
Alternatively, one finds the $\nu\rho$-components for the second choice as follows 
\begin{equation}
\begin{aligned}
\hat{Q} (\chi)&=\frac{1}{16\pi} \int_{\mbox{H}} dS\, \sqrt{\Omega} \bigg \{\sigma \Phi^2_0 \Big[2\kappa T-Y \theta \Big]+k \bigg [Y\frac{(2\kappa\hat{\lambda}+\theta^2-2 \theta^{'} )}{2\sqrt{\Omega}}+Y \frac{\theta \Omega^{'}}{2 \Omega^{3/2}}+Y^{'} \frac{\Omega^{'}}{2 \Omega^{3/2}}\\
&\hskip 3cm +Y \frac{\Omega^{'2}}{8 \Omega^{5/2}}+\lambda^{(0)}\frac{\Omega^{'}}{2 \Omega^{3/2}}-T \frac{\theta}{\sqrt{\Omega}}+Y\frac{(A^{(0)}_x)^2}{2 \sqrt{\Omega}}-3 \lambda^{(0)} \frac{A^{(0)}_x}{\sqrt{\Omega}}\bigg]\\
&\hskip 3cm+\beta \Phi^{-2}_0 \Big[\lambda^{(0)}A^{(1)}_\nu-Y A^{(0)}_x A^{(1)}_\nu\Big] \bigg\}.
\end{aligned}
\end{equation}
Accordingly, one can easily demonstrate that the variation of $\nu\rho$-component of the asymptotic charge via a second generator $\chi_2$ for these choices at the zeroth order respectively yields a closed algebra
\begin{equation}
\delta_{\chi_2} \hat{Q}^{(\nu\rho)} (\chi_1)=\hat{Q}^{(\nu\rho)}_{[\chi_1, \chi_2]}+\hat{{\cal C}}^{(\nu\rho)}(\chi_1, \chi_2),
\end{equation}
where the central extensions for these two choices turn out to be respectively as follows
\begin{equation}
\begin{aligned}
\hat{{\cal C}}^{(\nu\rho)}_{(A^{(0)}_x=0)}(\chi_1, \chi_2)&=\frac{1}{16\pi} \int_{\mbox{H}} dS\, \sqrt{\Omega} \bigg \{\sigma \Phi^2_0 \Big[2\kappa (Y_1\partial_x T_2 -T_1 \partial_x Y_2)\Big]\\
&\hskip 1.2 cm +k \bigg [Y_1 \partial_x Y_2\frac{(2\kappa\lambda+\theta^2-2 \theta^{'} )}{2\sqrt{\Omega}}+Y_1 \partial_x Y_2 \frac{\theta \Omega^{'}}{\Omega^{3/2}}+Y^{'}_1 \partial_x Y_2 \frac{\Omega^{'}}{2 \Omega^{3/2}}\\
&\hskip 1.2 cm +Y_1 \partial_x Y_2 \frac{\Omega^{'2}}{8 \Omega^{5/2}}+\lambda^{(0)}_1 \partial_x Y_2 \frac{\Omega^{'}}{2 \Omega^{3/2}}-T_1 \partial_x Y_2 \frac{\theta}{\sqrt{\Omega}} \\
& \hskip 1.2 cm-Y_1 \partial_x T_2 \frac{ \kappa \Omega^{'}}{\Omega^{3/2}}+  T_1 \partial_x T_2 \frac{2\kappa}{\sqrt{\Omega}} \bigg] \bigg\} ,
\label{centralext1}
\end{aligned}
\end{equation}
\begin{equation}
\begin{aligned}
\hat{{\cal C}}^{(\nu\rho)}_{(A^{(1)}_x=0)}(\chi_1, \chi_2)&=\frac{1}{16\pi} \int_{\mbox{H}} dS\, \sqrt{\Omega} \bigg \{\sigma \Phi^2_0 \Big[2\kappa (Y_1\partial_x T_2 -T_1 \partial_x Y_2) \Big]\\
&+k \bigg [Y_1 \partial_x Y_2 \frac{(2\kappa\lambda+\theta^2-2 \theta^{'} )}{2\sqrt{\Omega}}+Y_1 \partial_x Y_2 \frac{\theta \Omega^{'}}{\Omega^{3/2}}+Y^{'}_1 \partial_x Y_2 \frac{\Omega^{'}}{2 \Omega^{3/2}}\\
&\hskip 0.5 cm +Y_1 \partial_x Y_2 \frac{\Omega^{'2}}{8 \Omega^{5/2}}+\lambda^{(0)}_1 \partial_x Y_2 \frac{\Omega^{'}}{2 \Omega^{3/2}}-T_1 \partial_x Y_2 \frac{\theta}{\sqrt{\Omega}}\\
&\hskip 0.5 cm + Y_1 \partial_x Y_2 \frac{(A^{(0)}_x)^2}{ \sqrt{\Omega}}-3 [\lambda^{(0)}_1 \partial_x Y_2-\frac{1}{3}(Y_1 \partial_x \lambda^{(0)}_2)] \frac{A^{(0)}_x}{\sqrt{\Omega}} \\
&-Y_1 \partial_x T_2 \frac{\kappa\Omega^{'}}{\Omega^{3/2}}+T_1 \partial_x T_2 \frac{2\kappa }{\sqrt{\Omega}}-\lambda^{(0)}_1 \partial_x \lambda^{(0)}_2\frac{3}{\sqrt{\Omega}}\bigg]\\
&-\beta \Phi^{-2}_0 \Big[(Y_1 \partial_x Y_2) A^{(0)}_x A^{(1)}_\nu+(Y_1 \partial_x \lambda^{(0)}_2)A^{(1)}_\nu\Big] \bigg\}.
\label{centralext2}
\end{aligned}
\end{equation}
Observe that the central extensions in Eq.(\ref{centralext1}) and Eq.(\ref{centralext2}) associated to the different choices (i.e., $A^{(0)}_x=0$ or $A^{(1)}_x=0$) are \emph{spacetime point independent} once the integrals are performed over the horizon. Finally, let us notice that by mean of the Eq.(\ref{suprtrswitt}), one can easily split the components of the asymptotic charge into its supertranslation, superrotation and multiple-charge sectors as in \cite{SetareAdami5} for the first choice as follows
\begin{equation}
\begin{aligned}
{\cal T}_{(m, n)}&=\frac{1}{16\pi} \int_{\mbox{H}} dz d\bar{z}\, \sqrt{\gamma} \bigg[ 2\kappa \sigma \Phi^2_0-\frac{\theta}{\sqrt{\Omega}} \bigg]z^m \bar{z}^n .\\
{\cal Y}_m&=\frac{1}{16\pi} \int_{\mbox{H}} dz d\bar{z}\, \sqrt{\gamma}\bigg \{\sigma \Phi^2_0 \theta-k \Big[ \frac{(2\kappa\hat{\lambda}+\theta^2-2 \theta^{'} )}{2\sqrt{\Omega}}+ \frac{\theta \Omega^{'}}{2 \Omega^{3/2}}-\Big (\frac{\Omega^{'}}{2 \Omega^{3/2}}\Big )^{'}+\frac{\Omega^{'2}}{8 \Omega^{5/2}}\Big] \bigg\} z^{m+1}\\
{\cal \bar{Y}}_m&=\frac{1}{16\pi} \int_{\mbox{H}} dz d\bar{z}\, \sqrt{\gamma}\bigg \{\sigma \Phi^2_0 \theta-k \Big[ \frac{(2\kappa\hat{\lambda}+\theta^2-2 \theta^{'} )}{2\sqrt{\Omega}}+ \frac{\theta \Omega^{'}}{2 \Omega^{3/2}}-\Big (\frac{\Omega^{'}}{2 \Omega^{3/2}}\Big )^{'}+\frac{\Omega^{'2}}{8 \Omega^{5/2}}\Big] \bigg\} \bar{z}^{m+1}\\
{\cal Q}_{m,n}&=\frac{1}{16\pi} \int_{\mbox{H}} dz d\bar{z}\, \sqrt{\gamma} \Big [k\frac{\Omega^{'}}{2 \Omega^{3/2}}+\beta \Phi^{-2}_0 A^{(1)}_\nu  \Big] z^m \bar{z}^n,
\label{softmodes1}
\end{aligned}
\end{equation}
where $\gamma$ is the determinant of the induced metric associated to complex coordinates $z$ and $\bar{z}$. Similarly, one gets the relevant decomposition of the charge inducing asymptotic symmetries for the second choice as
\begin{equation}
\begin{aligned}
{\cal T}_{(m, n)}&=\frac{1}{16\pi} \int_{\mbox{H}} dz d\bar{z}\, \sqrt{\gamma} \bigg[ 2\kappa \sigma \Phi^2_0-\frac{\theta}{\sqrt{\Omega}} \bigg]z^m \bar{z}^n .\\
{\cal Y}_m&=\frac{1}{16\pi} \int_{\mbox{H}} dz d\bar{z}\, \sqrt{\gamma}\bigg \{\sigma \Phi^2_0 \theta-k \Big[ \frac{(2\kappa\hat{\lambda}+\theta^2-2 \theta^{'} )}{2\sqrt{\Omega}}+ \frac{\theta \Omega^{'}}{2 \Omega^{3/2}}-\Big (\frac{\Omega^{'}}{2 \Omega^{3/2}}\Big )^{'}+\frac{\Omega^{'2}}{8 \Omega^{5/2}}+\frac{(A^{(0)}_x)^2}{2 \sqrt{\Omega}}\Big] \\
& \hskip 3 cm-\beta \Phi^{-2}_0A^{(0)}_x A^{(1)}_\nu \bigg\} z^{m+1}\\
{\cal \bar{Y}}_m&=\frac{1}{16\pi} \int_{\mbox{H}} dz d\bar{z}\, \sqrt{\gamma}\bigg \{\sigma \Phi^2_0 \theta-k \Big[ \frac{(2\kappa\hat{\lambda}+\theta^2-2 \theta^{'} )}{2\sqrt{\Omega}}+ \frac{\theta \Omega^{'}}{2 \Omega^{3/2}}-\Big (\frac{\Omega^{'}}{2 \Omega^{3/2}}\Big )^{'}+\frac{\Omega^{'2}}{8 \Omega^{5/2}}+\frac{(A^{(0)}_x)^2}{2 \sqrt{\Omega}}\Big] \\
& \hskip 3 cm-\beta \Phi^{-2}_0A^{(0)}_x A^{(1)}_\nu \bigg\} \bar{z}^{m+1}\\
{\cal Q}_{m,n}&=\frac{1}{16\pi} \int_{\mbox{H}} dz d\bar{z}\, \sqrt{\gamma} \Big [ k\Big(\frac{\Omega^{'}}{2 \Omega^{3/2}}-3\frac{A^{(0)}_x}{\sqrt{\Omega}}\Big)+\beta \Phi^{-2}_0 A^{(1)}_\nu  \Big] z^m \bar{z}^n.
\label{softmodes2}
\end{aligned}
\end{equation}

\section{Conclusion}

The Weyl-invariant TMG is a ghost and tachyon-free (unitary) model that unifies the usual TMG, the Topologically Massive Electrodynamics and a conformally-coupled Proca mass term via the Weyl's gauge symmetry \cite{DengizErcanTekin}. It has been also demonstrated that the Weyl's gauge symmetry is spontaneously broken \'{a} la Higgs and Coleman-Weinberg mechanisms in (A)dS and flat vacua, respectively. Thus, the model brings a legitimate explanation to the origin of graviton mass through the symmetry-breaking mechanism as in the Standard Model Higgs mechanism. Additionally, the theory turns out to be unitary at the tree-level for certain regions of parameters. In this work, we would like to analyze the asymptotic structure of the Weyl-gauged TMG and reveal the associated algebra among the asymptotic charges. For this purpose, we have applied the \emph{modified} LW covariant phase space method which is introduced in \cite{Tachikawa} to cope with the Lagrangian possessing gravitational Chern-Simon term to the Weyl-gauged TMG. Accordingly, we have calculated the relevant quasi-local on-shell conserved charges associated to the enlarged symmetry $\chi=(\xi, \lambda)$ (that is, the diffeomorphism plus Weyl's $U(1)$ gauge symmetry). With the assumption of legitimate decay conditions for the existing propagating fields, we have subsequently evaluated the asymptotic charges generating asymptotic symmetries in the vicinity of a stationary black hole as in \cite{Tamburino, Giribet1, Giribet2, SetareAdami1, SetareAdami2, SetareAdami5}. From the field equations, we have found that the surface gravity $\kappa$ and $\Phi_{0}$ need to be constant, whereas $A^{(0)}_\nu$ should vanish. Since the gauge parameters are field dependent, we have defined an appropriate modified Lie derivative in consistent with the literature. With these modified brackets, we have demonstrated that the algebra among the enhanced asymptotic symmetry generators $\chi=\chi(T, Y, \lambda^{(0)})$ turns out to be closed.  For the particular choices of the central extensions, we have also shown that the asymptotic LW charge associated to asymptotic symmetry generators is integrable and the (modified) algebras among them are \emph{closed}. In general, the algebras among the modes consist of a family of supertranslation $(T_{(m, n)})$, two family of Witt algebras $(Y_m\,\mbox{and}\,\bar{Y}_m)$ and a family of multiple charges $(\lambda^{(0)}_{(m, n)})$. As a future direction, one can specifically study the black hole solutions of the model by using the conserved charges obtained in this article to get the energy, angular momentum and entropy of the black hole and investigate the first law of black hole thermodynamics. We intend to do this in future.

\section{Acknowledgments}

 We would like to thank Hamed Adami, Kamal Hajian, Gaston Giribet, Laura Donnay, Bayram Tekin, M.M. Sheikh-Jabbari for fruitful suggestions and discussions etc. The works of S.D. and E.K. are supported by the TUBITAK Grant No. 119F241.

\end{document}